\documentclass[rmp,aps,twocolumn]{revtex4}
\usepackage{graphicx,psfrag}
\usepackage{amsmath,amssymb}

\newcommand{\mycaption}[1]{\caption{#1}}
\newcommand{\figurecontents}[1]{\begin{center}#1\end{center}}

\begin{document}
\title{Elastic lever arm model for myosin V}
\author{Andrej Vilfan}
\affiliation{J. Stefan Institute, Jamova 39, 1000
  Ljubljana, Slovenia}
\altaffiliation[Part of the work has been conducted at ]{Max Planck Institute
  for the Physics of Complex Systems (MPIPKS), N{\"o}thnitzer Str. 38, 01187
  Dresden, Germany} \email{andrej.vilfan@ijs.si} \date{March 11, 2005}

\begin{abstract}
  We present a mechanochemical model for myosin V, a two-headed processive
  motor protein.  We derive the properties of a dimer from those of an
  individual head, which we model both with a 4-state cycle (detached, attached
  with ADP.Pi, attached with ADP and attached without nucleotide) and
  alternatively with a 5-state cycle (where the power stroke is not tightly
  coupled to the phosphate release).  In each state the lever arm leaves the
  head at a different, but fixed, angle.  The lever arm itself is described as
  an elastic rod.  The chemical cycles of both heads are coordinated
  exclusively by the mechanical connection between the two lever arms.  The
  model explains head coordination by showing that the lead head only binds to
  actin after the power stroke in the trail head and that it only undergoes its
  power stroke after the trail head unbinds from actin. Both models (4- and
  5-state) reproduce the observed hand-over-hand motion and fit the measured
  force-velocity relations.  The main difference between the two models
  concerns the load dependence of the run length, which is much weaker in the
  5-state model. We show how systematic processivity measurement under varying
  conditions could be used to distinguish between both models and to determine
  the kinetic parameters.
\end{abstract}

\keywords{}

\maketitle

\section*{Introduction}

Myosin V is a motor protein involved in different forms of intracellular
 transport \cite{Reck-Peterson.Mercer2000,Vale2003b}.  Because it was the first
 discovered processive motor from the myosin superfamily and due to its unique
 features, including a very long step size, it has drawn a lot of attention in
 recent years and now belongs to the best studied motor proteins.  The
 experiments have characterized it mechanically
 \cite{Mehta.Cheney1999,Rock.Spudich2000,Rief.Spudich2000,Veigel.Molloy2002,Purcell.Sweeney2002},
 biochemically
 \cite{De_La_Cruz.Sweeney1999,De_La_Cruz.Ostap2000a,De_La_Cruz.Ostap2000b,Yengo.Sweeney2002,Purcell.Sweeney2002},
 optically \cite{Ali.Ishiwata2002,Forkey.Goldman2003,Yildiz.Selvin2003} and
 structurally
 \cite{Walker.Knight2000,Burgess.Trinick2002,Wang.Sellers2003,Coureux.Houdusse2003}.
 These studies have shown that myosin V walks along actin filaments in a
 hand-over-hand fashion \cite{Yildiz.Selvin2003} with an average step size of
 about 35 nm, roughly corresponding to the periodicity of actin filaments
 \cite{Mehta.Cheney1999,Rief.Spudich2000,Veigel.Molloy2002,Ali.Ishiwata2002}, a
 stall force of around 2 pN \cite{Rief.Spudich2000} and a run length of a few
 microns \cite{Rief.Spudich2000,Sakamoto.Sellers2003,Baker.Warshaw2004}.  Under
 physiological conditions, ADP release was shown to be the time limiting step in
 the duty cycle \cite{De_La_Cruz.Sweeney1999,Rief.Spudich2000}.  Two stages of
 the power stroke have been resolved: one about 20nm, possibly connected with
 the release of phosphate, and another one of 5nm, probably occurring upon
 release of ADP \cite{Veigel.Molloy2002}.  Despite all this progress, the
 definite answer to the questions how the mechanical and the chemical cycle are
 coupled and how the heads communicate with each other to coordinate their
 activity has not yet been found.

 Theoretical models for processive molecular motors can follow different goals.
 What most models have in common is that they identify a few long-living states
 in the mechanochemical cycle and assume stochastic (Markovian) transitions
 between them.  The differences between models start in the way these states are
 chosen.  An approach that has been applied to myosin V
 \cite{Kolomeisky.Fisher2003}, kinesin
 \cite{peskin95,schief2001,Thomas.Tawada2002}, as well as to other biological
 mechanisms of force generation, including actin polymerization \cite{peskin93}
 and RNA polymerase \cite{wang-oster98}, models the motors as stochastic
 steppers.  These models describe the whole motor as an object that can go
 through a certain number of conformations (typically a few) with different
 positions along the track.  After the completion of one cycle (which is, in
 models for myosin V and kinesin, tightly coupled to the hydrolysis of one ATP
 molecule), the motor advances by one step.  All steps are reversible and at
 loads above the stall the motor is supposed to walk backwards and thereby
 regenerate ATP.  The approach has been particularly useful for interpreting the
 measured force-velocity relations and relating them to the kinetic parameters
 and positions of substeps
 \cite{schief2001,Fisher.Kolomeisky2001,Kolomeisky.Fisher2003}.  A limitation of
 such models is that they assume coordinated activity of both heads rather than
 explaining it.  They also assume that the motor strictly follows the regular
 cycle and there is no place for events like steps of variable length and
 dissociation from the track, although the latter can be incorporated into the
 models by proposing a different dissociation rate for each state in the cycle.

 In this Article we present a physical model for the processive motility of
 myosin V.  The basic building block of our model is an individual head, which
 we model in a similar way as the models for conventional myosins do
 \cite{hill74}, albeit with different rate constants.  The head is connected to
 the lever arm, which we model as an elastic rod, whose geometry we infer from
 electron microscopy studies \cite{Walker.Knight2000,Burgess.Trinick2002}.  The
 two lever arms are connected through a flexible joint and this is the exclusive
 way of communication between them.  We will derive the properties of the dimer
 from those of the individual head.

 \section*{The Model}

\begin{figure}
   \figurecontents{
     A)\includegraphics{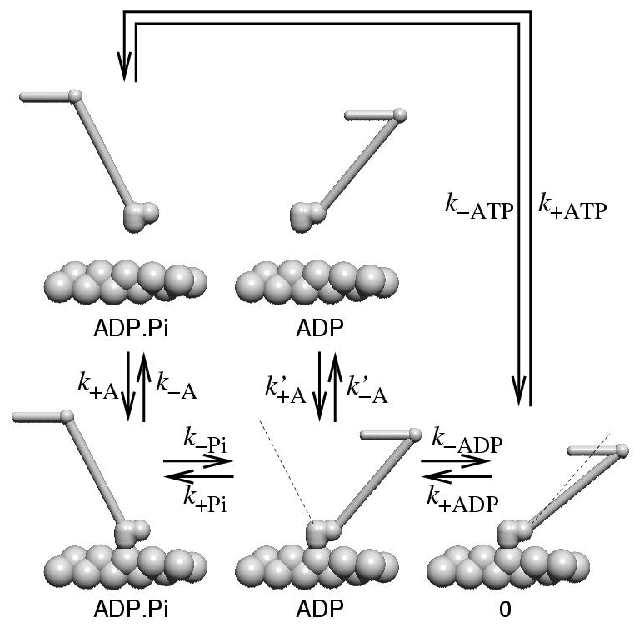}\\
     B)\includegraphics{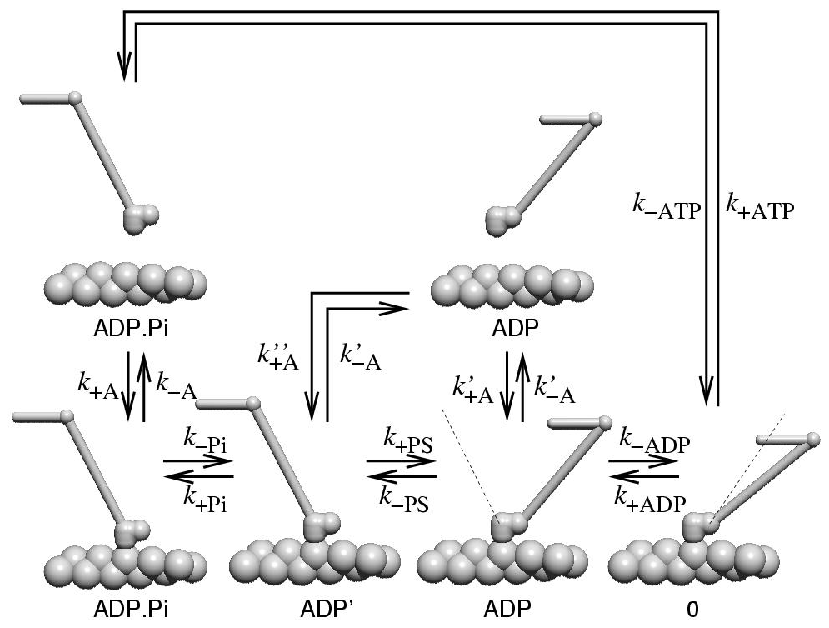}
   }
   \mycaption{A) The mechanochemical cycle of each individual head.  The head
     attaches to actin in the state with ADP and Pi bound on it, undergoes a
     large conformational change upon Pi release, another smaller conformational
     change upon ADP release, then binds ATP and enters the very weakly bound
     state, which dissociates quickly. B) The mechanochemical cycle in the
       5-state  model.  In this scenario, the phosphate release and the
     power stroke are two separate transitions.}
   \label{fig:2}
 \end{figure}

 To describe each myosin V head we use a model based on the 4-state cycle as
 postulated by \citet{lymn71} and used in many quantitative muscle models
 \cite{hill74} (Fig.~\ref{fig:2}A).   We restrict ourselves to
 the long-living states in the cycle: detached with ADP.Pi, bound with ADP.Pi,
 bound with ADP, detached with ADP and bound without a nucleotide. The bound
 state with ATP and the free state with ATP have both been found to be very
 short-lived \cite{De_La_Cruz.Sweeney1999} and we therefore omit them in our
 description, i.e., we assume that binding of ATP to a bound head leads to
 immediate detachment and ATP hydrolysis.  The detached state without a
 nucleotide is very unlikely to be occupied because of the low transition rates
 leading to it and we omit it from our scheme as well.

 One question that has not yet been definitely answered, is whether Pi release
 occurs before or during the power stroke, i.e., whether a head which is
 mechanically restrained form conducting its power stroke can release Pi or not.
 The 4-state model assumes a tight linkage between the Pi release and the power
 stroke. 
 While the 4-state model has been successfully applied to myosin II (e.g.,
 \citet{duke99,vilfan2003b}), recent experimental evidence suggests that the lead head can
 release Pi before the power stroke \cite{Rosenfeld.Sweeney2004}.  We
 therefore also discuss an alternative 5-state model. In the 5-state model we
 introduce an additional state ADP$'$ in which the phosphate is already
 released, but the lever-arm is still in the pre-powerstroke state.  The next
 transition, ADP release, however, is still linked to the completion of the full
 power-stroke. This is necessary in order to explain head coordination and also
 in agreement with experiments that show a strain-dependence in the ADP release
 rate in single-headed molecules \cite{Veigel.Molloy2002}.  The extended duty
 cycle of a head is shown in Fig.~\ref{fig:2}B. 

 \begin{figure}[htbp]
   \figurecontents{
     \includegraphics{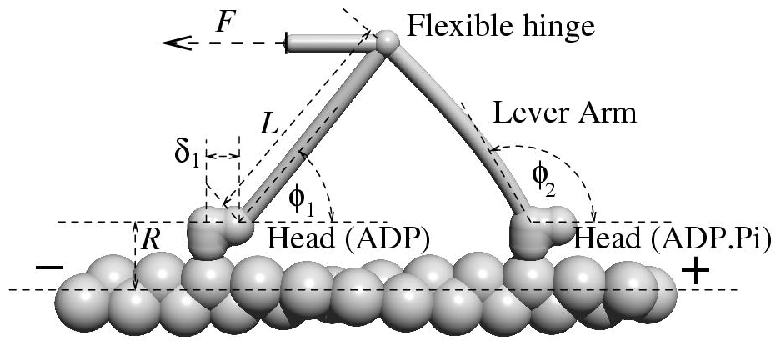}
   }
   \mycaption{The myosin V dimer is modeled as two heads, each connected to a
     lever arm which leaves the head at a certain angle $\phi$, depending on the
     state of the head.  The two lever arms, modeled as elastic beams, are
     connected with a flexible joint, which is also connected to the external
     load.}
   \label{fig:1}
 \end{figure}

 \begin{table}
   \mycaption{Geometric parameters of a myosin V head (see also
     Fig.~\ref{fig:1} for their definition).}
   \label{tab:I}
   \begin{center}
     \begin{tabular}{lll}
       \hline
       Lever arm length & $L$ & $26\,{\rm nm}$\\
       Lever arm  start   & $R$ & $8\,{\rm nm}$\\
       Lever arm  start   & $\delta_{\rm ADP.Pi}$ & $0\,{\rm nm}$\\
       Lever arm  start   & $\delta_{\rm ADP,apo}$ & $3.5\,{\rm nm}$\\
       Angle ADP.Pi & $\phi_{\rm ADP.Pi}$ & $115^{\circ}$\\
       Angle ADP & $\phi_{\rm ADP}$ & $50^{\circ}$\\
       Angle apo & $\phi_{\rm apo}$ & $40^{\circ}$\\
       \hline
     \end{tabular}
   \end{center}
 \end{table}

 A head always binds to an actin subunit in the same relative position.  In each
 state, the proximal end of the lever arm leaves the head in a fixed direction
 in space, determined by the polar angle $\phi$ towards the filament plus end
 and the azimuthal angle $\theta=\theta_0 i$ of the actin subunit $i$ to which
 the head is bound.  The geometry of the molecule and the angles were inferred
 from images obtained with electron microscopy
 \cite{Walker.Knight2000,Burgess.Trinick2002}.  They are summarized in Table
 \ref{tab:I}.  In our calculations we assume a 13/6 periodicity of the
 actin helix (6 rotations per 13 subunits), which means $\theta_0=2\pi \times
 6/13$.

\begin{figure*}[htbp]
 \figurecontents{
   \includegraphics{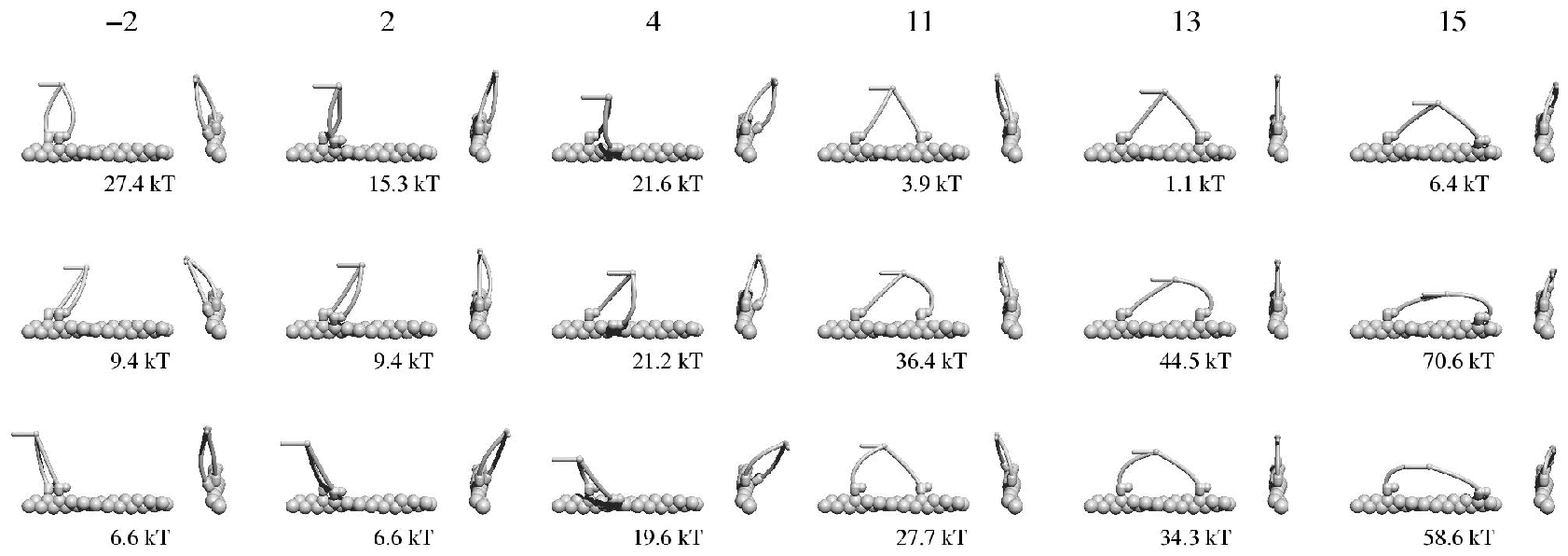}
 }
 \mycaption{Calculated shapes and bending energies of dimers, bound $i$
   subunits apart ($i=-2,2,\ldots,15$) and in different states: first in post-,
   second in the pre-powerstroke state (upper row), both in the post-powerstroke
   state (middle row) and both in the pre-powerstroke state (bottom row).  Each
   configuration is shown in side and front view. If both heads are in the same
   state (bottom two rows) there is a significant cost in elastic energy needed
   to buckle one of the lever arms.  Binding of the lead head before the trail
   head undergoes the power stroke is therefore unlikely.  }
 \label{fig:3}
 \end{figure*}

 We assume that the lever arm has the properties of a linear, uniform and
 isotropic elastic rod, described with the bending modulus $EI$.  Then the local
 curvature $\kappa$ is determined from $M=EI {\mathbf\kappa}$, where $M$ is the
 local bending moment (torque).  The lever arms from both heads are joined
 together (and to the tail) with a flexible joint which allows free rotation in
 all directions. For a certain configuration of chemical states, binding
 sites of both heads and a given external force, the three-dimensional shape and
 the bending energy of both lever arms can be calculated numerically as
 described in the Appendix.  Some of the calculated shapes are shown in
 Fig.~\ref{fig:3}. 

 We calculate the free energy of a dimer state as
 \begin{equation}
 G=G_1 + G_2 + U_1 + U_2 + F x\;,
 \label{eq:1}
 \end{equation}
 where $G_1$ and $G_2$ are the intrinsic free energies of both heads (which
 depend on the chemical state of the head and the concentrations of
 nucleotides), $U_1$ and $U_2$ are the energies stored in the elastic
 deformation of each lever arm, and $F x$ is the work done against the external
 load ($x$ denotes the coordinate of the flexible joint along the filament axis
 with positive values towards the plus end, while positive values of $F$ denote
 a force pulling towards the minus end, against the direction of motion of an
 unloaded motor).

 \subsection*{Transition rates}

 There are two exact statements we can make about the kinetic rates of the duty
 cycle that follow from the principle of detailed balance.  The first statement
 relates the forward and the backward rate of any reaction to the free energy
 difference between the initial and the final state.  For any transition the
 principle of detailed balance states that
 \begin{equation}
   \label{eq:2}
   \frac{k_{+i}}{k_{-i}} =
   \frac{k^0_{+i}}{k^0_{-i}}  e^{-\frac{\Delta U + F \Delta x}{k_B T}}
 \end{equation}
 where $\Delta U$ denotes the change in elastic energy of the dimer 
 and $F\Delta x$ the work performed against the external load.  

 The second exact statement can be derived by multiplying together the detailed
 balance conditions for a monomer in the absence of any external force along a
 closed pathway in Fig.~\ref{fig:2}. After one cycle the free energy of the
 bound monomeric head returns to its initial value, while the total free energy
 change in the system equals the amount gained from the hydrolysis of one ATP
 molecule.  The resulting relation reads
 \begin{multline}
   \label{eq:3}
   \frac{k^0_{\rm +A} k^0_{\rm -Pi} k^0_{\rm -ADP} k_{\rm +ATP} [{\rm ATP}]}{k_{\rm -A} k^0_{\rm +Pi} [{\rm Pi}] k^0_{\rm +ADP}
   [{\rm ADP}] k^0_{\rm -ATP}} \\
 =e^{\frac{\Delta G_{ATP}}{k_B T}}=e^{\frac{\Delta G^0}{k_B T}}
   \frac{[{\rm ATP}]}{[{\rm ADP}][{\rm Pi}]}
 \end{multline}
 and provides an important constraint on the kinetic rates of the model. 
 In the 5-state model, we obtain an equivalent equation,
 \begin{multline}
   \label{eq:4}
   \frac{k^0_{\rm +A} k^0_{\rm -Pi} k^0_{\rm +PS} k^0_{\rm -ADP} k_{\rm +ATP}
     [{\rm ATP}]}{k_{\rm -A} k^0_{\rm +Pi} [{\rm Pi}] k^0_{\rm -PS} k^0_{\rm
       +ADP}
     [{\rm ADP}] k^0_{\rm -ATP}} \\
   =e^{\frac{\Delta G^0}{k_B T}} \frac{[{\rm ATP}]}{[{\rm ADP}][{\rm Pi}]}\;.
 \end{multline}
 A similar statement also holds for the rates along the inner loop in the
 reaction scheme, which involves attachment, power stroke and detachment, all in
 the ADP state.  Because we assume that the detachment rate in the
 pre-powerstroke and the post-powerstroke state are both the same ($k'_{\rm
   -A}$), the relation reads
 \begin{equation}
   \label{eq:5}
 \frac{  k^{0\prime\prime}_{\rm +A} k^0_{\rm +PS} } {  k^{0\prime}_{\rm +A}  k^0_{\rm -PS} }=1 \;.
 \end{equation}

 When it comes to the actual force dependence of transition rates we have to
 rely on approximations. An approach that is most widely used when modeling
 motor proteins, but also other conformational changes, like the gating of ion
 channels, involves the Arrhenius theory of reaction rates \cite{hill74}.  It
 proposes that the protein has to reach an activation point ($x_a$) somewhere
 between the initial ($x_i$) and the final state ($x_f$) by thermal diffusion,
 but completes the reaction rapidly after that.  Therefore, the force dependence
 of the forward rate can be modeled as
 \begin{equation}
   \label{eq:6}
   k_{+i}=k_{+i}^0 e^{-\frac {U(x_a)-U(x_i)}{k_B T}} \qquad
   k_{-i}=k_{-i}^0 e^{-\frac {U(x_a)-U(x_f)}{k_B T}} 
 \end{equation}
 where $U(x)$ means the total potential (bending of both lever-arms and work
 done against the external load) which a head has to overcome to bring the lever
 arm angle into a given state. We use the variable $\epsilon$ to denote the
 relative position of the activation point between the initial and the final
 state, so that $x_a=(1-\epsilon) x_i +\epsilon x_f$.  Unless otherwise noted,
 we will assume $\epsilon=0.5$. Not precisely identical, but useful for
 practical purposes is also the approximation
 $U(x_a)=(1-\epsilon)U(x_i)+\epsilon U(x_f)$.  Therefore we get the following
 expression for the force-dependence of the transition rate:
 \begin{equation}
   \label{eq:7}
   k_{+i}=k_{+i}^0 e^{\frac{\epsilon \Delta U}{k_B T}}
 \end{equation}

 For reactions that involve the binding and unbinding of a head, Eq.~\ref{eq:2}
 is valid, but one expects the activation point to be much closer to the bound
 state.  The strain-dependence of the detachment rate for heads in the ADP and
 ATP.Pi state has not yet been measured and we therefore neglect it, assuming
 that the detachment rate is force-independent, $k_{\rm -A}\equiv k_{\rm -A}^0$.
 The attachment rate then relates to the potential difference as
 \begin{equation}
    k_{\rm +A}=k_{\rm +A}^0 e^{-\frac{\Delta U}{k_B T}}\;.
 \end{equation}

 \subsection*{Choice of kinetic parameters}

 \begin{table*}[htbp]
   \mycaption{Kinetic parameters of the model}
   \label{tab:II}
   \figurecontents{
     \begin{tabular}{lp{5.5cm}llp{6cm}}
       \multicolumn{2}{c}{Parameter} & \multicolumn{2}{c}{Value}& Source\\
       &&4-state&5-state&\\
       \hline\\
       $k^0_{\rm +A}$ & actin binding with ADP.Pi & $5000\,{\rm s^{-1}}$&
       $5000\,{\rm s^{-1}}$&est.~from run length \\
       $k_{\rm -A}$ & actin release with ADP.Pi & $1\,{\rm s^{-1}}$& $50\,{\rm
       s^{-1}}$&est.~from run length  \\
       $k^{0\prime}_{\rm +A}$ & actin binding with ADP & $5000\,{\rm s^{-1}}$& $5000\,{\rm s^{-1}}$&
       $\approx k^0_{\rm +A}$ \cite{De_La_Cruz.Sweeney1999} \\
       $k^{\prime}_{\rm -A}$ & actin release with ADP & $0.1\,{\rm s^{-1}}$ & $0.1\,{\rm s^{-1}}$ &  $0.032\,{\rm
         s^{-1}}$ \cite{De_La_Cruz.Sweeney1999}, $1.1 \,{\rm s^{-1}}$ \cite{Baker.Warshaw2004}\\
       $k^0_{\rm -Pi}$ & Pi release & $200\,{\rm s}^{-1}$ & $200\,{\rm s}^{-1}$
       & $>250\,{\rm s}^{-1}$ \cite{De_La_Cruz.Sweeney1999}, $110\,{\rm s}^{-1}$
       \cite{Yengo.Sweeney2004}, $228\,{\rm s}^{-1}$
       \cite{Rosenfeld.Sweeney2004}\\
       $\epsilon_{\rm -Pi}$ & activation point & $0.3$ &--&F-v relation at high loads\\
       $k^0_{\rm +Pi}$ & Pi binding & $10^{-4}\,{\mu \rm M}^{-1}{\rm s}^{-1}$ &
       $10^{-2}\,{\mu \rm M}^{-1}{\rm s}^{-1}$ &
       guess \\
       $k^0_{\rm +PS}$ & power stroke & -- &
       $10^{4}\,{\rm s}^{-1}$ &
       guess \\
       $k^0_{\rm -PS}$ & reverse stroke & -- &
       $0.05\,{\rm s}^{-1}$ &
       $k^0_{\rm +PS}/k^0_{\rm -PS}$ from the stall force \\
       $k^0_{\rm -ADP}$ & ADP release & $20\,{\rm s}^{-1}$ & $20\,{\rm s}^{-1}$ &
       $k_{\rm -ADP}=13\,{\rm s}^{-1}$ for dimers \cite{Rief.Spudich2000}\\
       $k^0_{\rm +ADP}$ & ADP binding & $12\,{\mu \rm M}^{-1}{\rm s}^{-1}$ & $12\,{\mu \rm M}^{-1}{\rm s}^{-1}$ &
       $12.6\,{\mu \rm M}^{-1}{\rm s}^{-1}$  \cite{De_La_Cruz.Sweeney1999},
       $14\,{\mu \rm M}^{-1}{\rm s}^{-1}$ \cite{Wang.Sellers2000}\\
       $k_{\rm +ATP}$ & ATP binding, actin release & $0.7 \,{\mu \rm M}^{-1}{\rm
       s}^{-1}$ & $0.7 \,{\mu \rm M}^{-1}{\rm s}^{-1}$ &
        $0.9 \,{\mu \rm M}^{-1}{\rm s}^{-1}$
       \cite{De_La_Cruz.Sweeney1999,Rief.Spudich2000},  $0.6-1.5 \,{\mu \rm
       M}^{-1}{\rm s}^{-1}$ \cite{Veigel.Molloy2002}\\
       $k^0_{\rm -ATP}$ & actin binding with ATP release & $0.07\,{\rm s}^{-1}$
       & $1.2\,{\rm s}^{-1}$ &
       Eq.~\ref{eq:3}, Eq.~\ref{eq:4}  \\
     \end{tabular}
   }
 \end{table*}

 Some of the transition rates in the cycle are well known from the literature.
 $k_{\rm -ADP}$ is the limiting rate both for running myosin V molecules and for
 single-headed constructs at low ATP concentrations.  The measured values are
 $13\,{\rm s}^{-1}$ \cite{Rief.Spudich2000} for dimers and $12\,{\rm s}^{-1}$
 \cite{De_La_Cruz.Sweeney1999}, $13$--$22\,{\rm s^{-1}}$
 \cite{Trybus.Freyzon1999}, and $4.5$--$7\,{\rm s}^{-1}$
 \cite{Molloy.Veigel2003} for monomers.  Because the actual rate in a dimer is
 slowed down as compared to the monomer, we use the value $k^0_{\rm
   -ADP}=20\,{\rm s}^{-1}$. The reverse rate, $k_{\rm +ADP}$ can be determined
 from the inhibitory effect of ADP on the velocity and has been estimated as
 $12.6\,{\rm \mu M}^{-1}{\rm s}^{-1}$ \cite{De_La_Cruz.Sweeney1999}, $4.5\,{\rm
   \mu M}^{-1}{\rm s}^{-1}$ \cite{Rief.Spudich2000}, $14\,{\rm \mu M}^{-1}{\rm
   s}^{-1}$ \cite{Wang.Sellers2000}.

 Equally well known is the rate for ATP binding, $k_{\rm +ATP}$, which has been
 measured as $0.9\,{\rm \mu M}^{-1}{\rm s}^{-1}$
 \cite{De_La_Cruz.Sweeney1999,Rief.Spudich2000}, $0.6$--$1.5\,{\rm \mu
   M}^{-1}{\rm s}^{-1}$ \cite{Veigel.Molloy2002}.  For the Pi release rate the
 estimates range from $k_{\rm -Pi}>250\,{\rm s}^{-1}$
 \cite{De_La_Cruz.Sweeney1999} to $110\,{\rm s}^{-1}$ \cite{Yengo.Sweeney2004}.
 We therefore use the value $k_{\rm -Pi}=200\,{\rm s}^{-1}$.

 There is some more discrepancy between the current values for the release rate
 from actin in the ADP state.  While direct measurements gave $k'_{-\rm
   A}=0.032\,{\rm s}^{-1}$ \cite{De_La_Cruz.Sweeney1999} and $0.08\,{\rm
   s}^{-1}$ \cite{Yengo.Sweeney2004}, a recent estimate from the run length led
 to a higher value of $1.1\,{\rm s}^{-1}$ \cite{Baker.Warshaw2004}. We use an
 intermediate value of $k'_{-\rm A}=0.1\,{\rm s}^{-1}$.  For the attachment rate
 in the ADP state, we set $k^{0 \prime}_{+\rm A}\approx k^0_{+\rm A}$, based on
 kinetic measurements \cite{De_La_Cruz.Sweeney1999}.

 This leaves us with a total of 4 unknown kinetic rates, of which 3 need to be
 estimated from the measured stepping behavior and run length data, while one
 can be determined from Eq.~\ref{eq:3}.

 \section*{Results}

 \subsection*{Choice of the value for the bending modulus}

 There are two ways to estimate the bending stiffness of the myosin V lever arm
 - one from its structure and analogy with similar molecules and the other one
 from the observed behavior of the dimeric molecule. The lever arm consists of 6
 IQ motifs, forming an $\alpha$-helix, surrounded by 6 calmodulin or other light
 chains \cite{Wang.Sellers2003,Terrak.Dominguez2003}.  One possible estimate for
 the stiffness of the lever arm can be obtained by approximating it with a
 coiled-coil domain, as has been done by \citet{Howard.Spudich1996}.  Generally,
 the stiffness of a semiflexible molecule is related to its persistence length
 $\ell_p$ as $EI=\ell_p k_B T$.  Howard and Spudich estimated the persistence
 length of a coiled-coil domain as $100\,{\rm nm}$, which yields $EI\approx
 400\,{\rm pN\,nm}^2$. Other researchers report values of $\ell_p=130\,{\rm nm}$
 for myosin \cite{Hvidt.Ferry1982} and $\ell_p=150\,{\rm nm}$ for tropomyosin
 \cite{Swenson.Stellwagen1989,Phillips.Chacko1996}.

\begin{figure}[htbp]
 \figurecontents{
   \includegraphics{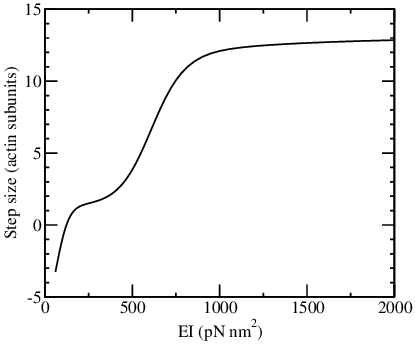}
 }
 \mycaption{The average step size under a load of $F=1.8\,{\rm pN}$ as a
   function of the lever arm elasticity $EI$. The step size was calculated from
   attachment probabilities of the lead head (ADP.Pi state) relative to the
   bound trail head (ADP state).}
   \label{fig:4}
 \end{figure}

 On the other hand, we can estimate the stiffness from the force a lever arm has
 to bear under conditions close to stall.  We do this by calculating the
 distribution of binding probabilities to different sites at $F=1.8\,{\rm pN}$,
 which is close to stall force.  We assume that the binding rate to each site is
 proportional to its Boltzmann weight, $\exp(- G / k_B T)$, which is equivalent
 to assuming that the activation point of the binding process is close to the
 final state and that the reverse reaction (detachment in the state with ADP.Pi)
 has no force-dependence in its rate.  The expectation value of the binding
 position of the lead head relative to the trail head is shown in
 Fig.~\ref{fig:4}.  It shows that a stiffness of $EI\gtrsim 1000\,{\rm pN\, nm^2
 }$ is necessary to allow stepping at loads of this magnitude.

  For these reasons, we use the value $EI=1500\,{\rm pN\,nm^2}$.  This
 corresponds to an elastic constant (measured at the joint) of
 \begin{equation}
   \label{eq:9}
   k=3 EI/L^3=0.25\,{\rm pN/nm}\;.
 \end{equation}
 The elastic constant for longitudinal forces (with respect to the lever arm) is
 much higher. If we approximate the lever arm with a homogeneous cylinder of
 radius $r=1\,{\rm nm}$, we can estimate it as $k_L=4 EI /(r^2 L)=230\,{\rm
   pN/nm}$.  We therefore neglect the longitudinal extensibility of the lever
 arm in all calculations.

 A similar value ($EI=1300\,{\rm pN\,nm^2}$) has also been obtained by analyzing
 data from optical trap experiments on single-headed myosin V molecules with
 different lever arm lengths \cite{Moore.Warshaw2004}. Even though it is
 somewhat larger (about 3 times) than the values estimated for myosin II
 \citep{Howard.Spudich1996}, there is no solid evidence that the structures with
 different light chains have the same bending stiffness.  On the other hand,
 there could have been some evolutionary pressure to increase the lever arm
 stiffness, as it is directly related to the stall force of myosin V.  While we
 are not able to give a definite answer to the question whether the lever arm
 behaves like a uniform elastic rod or whether there is a pliant region close to
 the head, we favor the first hypothesis because the estimated lever arm
 elasticity already is more than sufficient to explain the mechanical properties
 of the dimeric molecule.

 \subsection*{Step size distribution}

 Figure \ref{fig:3} shows the energies stored in the elastic distortions of the
 lever arms of both heads in the pre-powerstroke or the post-powerstroke state.
 For example, if the first head is in the ADP.Pi state and the second head binds
 before the first one undergoes a power stroke, this is connected with an energy
 cost of $6.6\,k_B T$.  The attachment rate of the lead head before the power
 stroke in the trail head is therefore more than 100 times slower than after the
 power stroke.

\begin{figure}[htbp]
   \figurecontents{
     \includegraphics{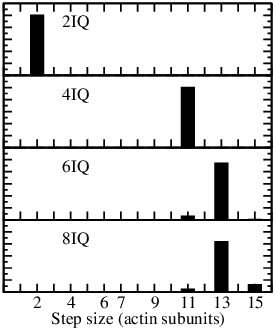}
   }
   \mycaption{Step size distribution for 4 different lever arm lengths L: 10nm
     (2IQ), 18nm (4IQ), 26nm (6IQ) and 34nm (8IQ) and no external load.  The
     histograms show the probability that a lead head (ADP.Pi state) will bind
     $i$ sites in front of the trail head in the post-powerstroke ADP state. The
     probabilities were determined from the Boltzmann factors, resulting from
     the elastic distortion energy of the configuration.  Azimuthal distortion
     plays a crucial role role in determining the step size, which is the reason
     why the binding is always concentrated on sites 2, 11, 13 and 15.  Taking
     into account the fluctuations in the actin would lead to a broader
     distribution, in better agreement with experiments
     \citep{Walker.Knight2000}.}
   \label{fig:5}
 \end{figure}

 Because the lead head normally attaches to actin while the trail head is in the
 ADP state, we can determine the probability that the lead head binds to an
 actin site $i$ subunits in front of the trail head from the Boltzmann factors
 formed from the bending energy in the final configuration, $P_i \propto \exp (-
 (U_1+U_2)/k_B T)$.  Here $U_1+U_2$ denotes the sum of elastic energies stored
 in both lever arms if the trail head is in the ADP state and the lead head in
 the ADP.Pi state, bound $i$ sites in front of the trail head.  The resulting
 distributions for different lever arm lengths are shown in Fig.~\ref{fig:5}.
 For the lever arm consisting of 6 IQ motifs, the result is a mixture of 11 and
 13 subunit steps, whereby 13 subunits dominate.  Azimuthal distortion plays a
 major role in the bending energy, therefore binding is only likely to sites 2,
 11, 13 and 15, on which the azimuthal angles of both heads differ by not more
 than $27^\circ$.

 \subsection*{The gated step in the cycle}
 \label{sec:}

 A question that has been a subject of intense discussion is which step in the
 cycle is deciding for the coordination of the two heads.  A currently often
 favored hypothesis proposes that the lead head undergoes its power stroke
 immediately after binding, thereby storing energy into elastic deformation of
 its lever arm and releasing it after the unbinding of the trail head.  An
 alternative hypothesis proposes that the release of the rear head is necessary
 for the power stroke in the front head.  As we will show below, our model
 favors this picture.  In the 4-state scenario, this implies that the lead head
 is waiting in the ADP.Pi.  In the 5-state scenario it is in the ADP$'$ state
 (the pre-powerstroke ADP state).  The trail head spends most of its cycle in
 the ADP state in both scenarios at saturating ADP concentrations.

 Because this model challenges the currently prevailing view, we should first
 critically review the arguments supporting it.  One argument includes the
 direct observation of telemark-shaped molecules, with the leading head leaning
 forward and then the lever arm tilted strongly backwards
 \cite{Walker.Knight2000}.  A more detailed image analysis, however, showed that
 the converter of the leading head is in the pre-powerstroke state
 \cite{Burgess.Trinick2002}.  Another piece of evidence comes from experiments
 by \citet{Forkey.Goldman2003} which show a fraction of tags on the lever arm
 (30-50\%) that do not tilt while moving, but again the data provide no
 conclusive proof because the method does not allow detection of tilts symmetric
 with respect to the vertical axis.  To conclude, one cannot say that the
 present experimental evidence excludes any of the two hypotheses about the
 moment of phosphate release and of the power stroke.

\begin{figure}[htbp]
   \figurecontents{
     \includegraphics{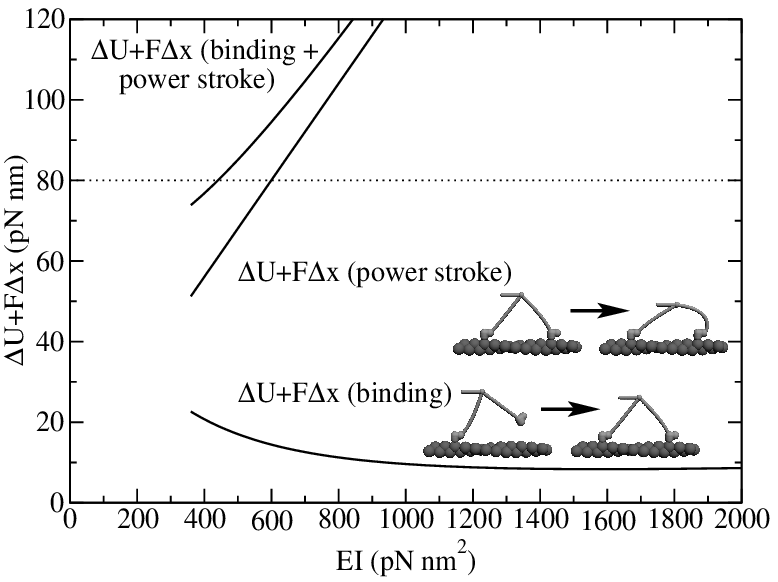}
   }
   \mycaption{The amount of energy needed for the binding of the lead head and
     the subsequent power-stroke, plotted against the lever arm elasticity.  The
     load pulling on the tail is $F=1.8\,{\rm pN}$. The lower curve shows the
     energy needed to pull the external load and distort the lever arms in order
     to bind the new lead head 13 sites in front of the trailing head.  Note
     that most of this work will be performed by Brownian motion, but the
     potential well in the bound state still has to be strong enough to
     stabilize the bound state. The middle curve shows the energy needed mainly
     for the distortion of the lever arms when the lead head undergoes a
     power-stroke before the trailing head detaches.  Since the sum of both
     cannot be higher than $80\,{\rm pN\,nm}$, we estimate that this
     hypothetical scenario would only be possible if the lever arm stiffness was
     $EI \lesssim 450\,{\rm pN\,nm^2}$.  This is inconsistent with other
     requirements of the model, so we rule this scenario out.}
   \label{fig:6}
 \end{figure}

 From the theoretical side, we will argue that in a model with linear elasticity
 the mechanism with immediate power stroke in the lead head cannot work under
 loads for which the motor is known to be operational.  It is known that the
 monomeric constructs of myosin V undergo a normal duty cycle
 \cite{De_La_Cruz.Sweeney1999,Yengo.Sweeney2002}, which means that no step in
 the cycle requires mechanical work from the outside for its completion (which
 would be, for example, the case if the head needed to be pulled away from actin
 to complete the cycle).  This excludes the possibility that the free energy
 gain connected with binding and the power stroke exceeds $\Delta G_{\rm
   ATP}=100\,{\rm pN nm}$, the total available energy for one cycle.  Because
 this and other transitions in the cycle need to be forward-running, we use the
 still conservative estimate that the free energy gain from binding and the
 power stroke cannot exceed $80\,{\rm pN nm}$.  On the other hand, we can
 estimate the free energy that would be necessary for a head to bind to a site
 13 units ahead and then undergo a conformational change.  The amount of energy
 needed to bring the dimer into the hypothetical state with both heads in the
 post-powerstroke state and a strong distortion, especially of the leading lever
 arm, is plotted in Fig.~\ref{fig:6}.  The calculation shows that
 the binding of the front head with the subsequent power stroke before the rear
 head detaches (for a load of $F=1.8\,{\rm pN}$) is only possible for values of
 $EI\lesssim 450\,{\rm pN\,nm^2}$, which is inconsistent with the lower estimate
 based on the observed step size (Fig.~\ref{fig:4}).  Of course, we
 cannot rule out that there is some additional state in the middle of the power
 stroke which  is occupied  immediately while the lead head waits for the trail
 head to detach.  But within the scope of the geometrical model with a single
 power stroke connected with the Pi release, we consider the scenario where the
 lead head instantaneously undergoes the power stroke without waiting for the
 detachment of the trail head unrealistic.

 \subsection*{Hidden power strokes in the dimer configuration}

 \begin{figure}[htbp]
   \figurecontents{
     \includegraphics{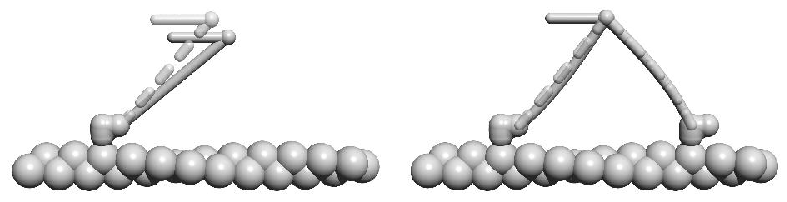}
   }
   \mycaption{For a single head, the $x$-component of the power-stroke upon ADP
     release equals 3.3nm (for zero load).  In the dimer with both heads bound,
     only 0.07nm of that power stroke reach the load.  As a consequence, the
     load-dependence of transition rates between states with both heads bound is
     negligible.}
 \label{fig:7}
 \end{figure}

 An immediate consequence of the elastic lever arm model is that the tail
 position is mainly determined by the geometry of the triangle and less by the
 conformations of individual heads.  For a monomeric head or a dimer bound by a
 single head, the power-stroke upon ADP release has an $x$-component (in the
 direction of the actin filament) of about $3.3\,{\rm nm}$ (Fig.~\ref{fig:7}).
 If the lead head is attached, however, the power stroke as measured on the tail
 is reduced by about a factor of 50.  The tail movement is also closely related
 to the force-dependence of transition rates, which means that transitions
 between states with both heads bound do not show any significant load
 dependence.  In the kinetic scheme we use here this implies that the rates of
 ADP release and ATP binding (the two rate limiting steps at low or forward
 loads) are both constant, in agreement with the flat F-v curve measured by
 \citet{Mehta.Cheney1999}.

 \begin{figure*}[htbp]
   \figurecontents{
     \begin{tabular}{ll}
       \hspace*{-0.5cm}\includegraphics{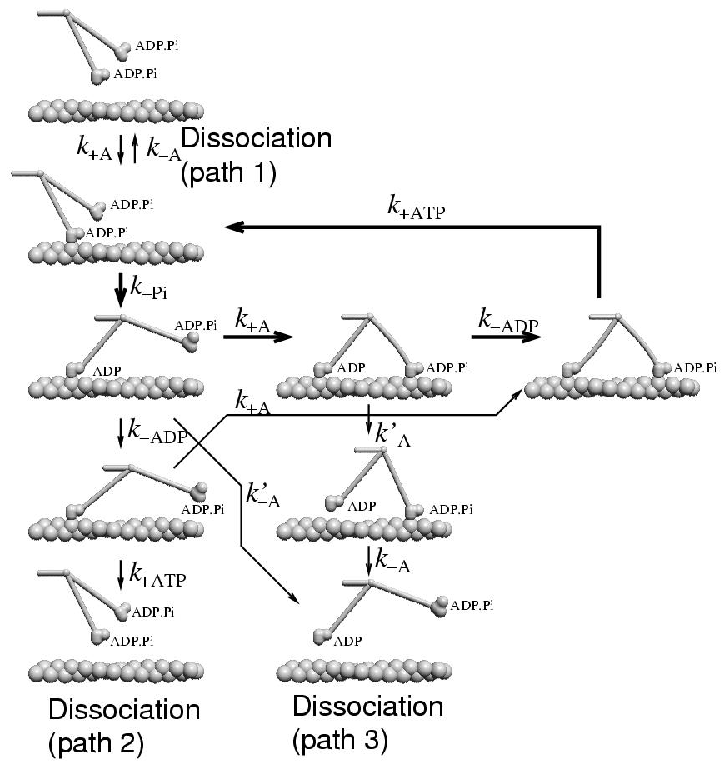}&    \includegraphics{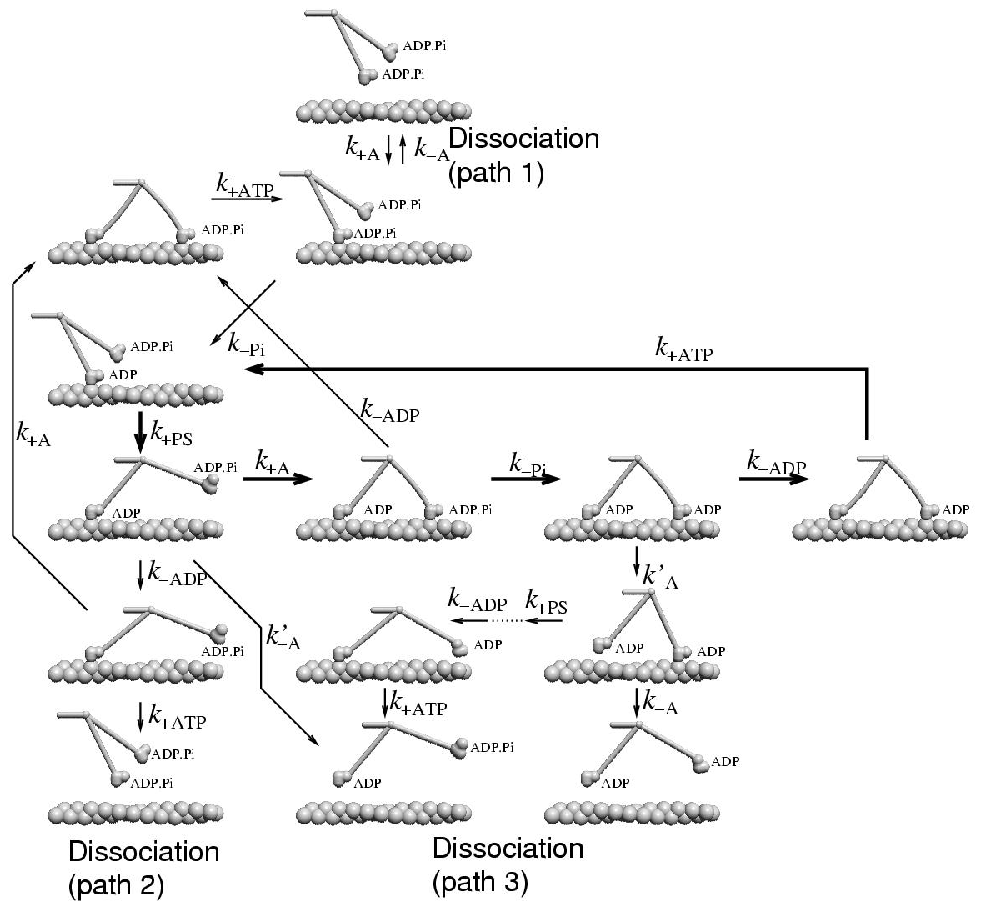}
 \\      \hspace*{-0.5cm}A)&B)
     \end{tabular}
   }
   \mycaption{ Most probable kinetic pathways for a dimer in the 4-state (A) and
     in the 5-state model (B).  The thick arrows denote the regular pathway and
     the thin arrows side branches that can result in dissociation from actin.
     Note that the simulation was not restricted to the pathways shown here, but
     included all possible combinations of transitions between monomer
   states.  }
   \label{fig:8}
 \end{figure*}

 \begin{figure}[htbp]
   \figurecontents{
 A)    \includegraphics{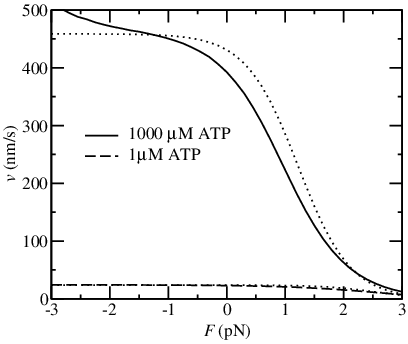}\\

 B)    \includegraphics{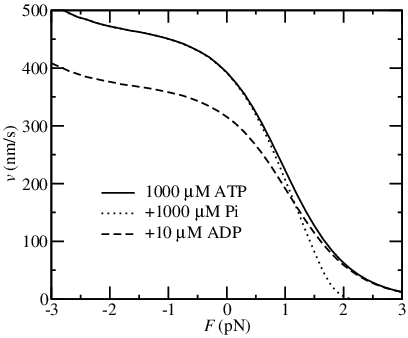}\\
   }
   \mycaption{A) 
     Force-velocity curves in the 4-state model, obtained from a stochastic
     simulation.  The solid curve shows the values for $1000\mu \rm M$ ATP and
     the dashed curve for $1 \mu \rm M$ ATP.  Both curves are compared with the
     prediction of the simplified analytical expression (Eq.~\ref{eq:13}),
     dotted lines.  The minor deviation is mainly due to cycles taking other
     pathways, neglected force-dependence of the ADP release rate and variation
     in the step size.  Note that the velocities above $\sim 2.5\,{\rm pN}$ are
     not well defined because the dissociation time becomes comparable with the
     step time.  B) Inhibition by ADP and Pi.  The force-velocity relation with
     1mM ATP is shown by the continuous line.  The dashed line shows the same
     relation with additional $10\mu{\rm M}$ ADP and the dotted line with 1mM
     phosphate.  The velocity reduction through ADP occurs at low or negative
     loads, while the inhibition by Pi only becomes significant close to stall
     conditions. }
   \label{fig:9}
 \end{figure}

 \begin{figure}[htbp]
   \figurecontents{
     \includegraphics{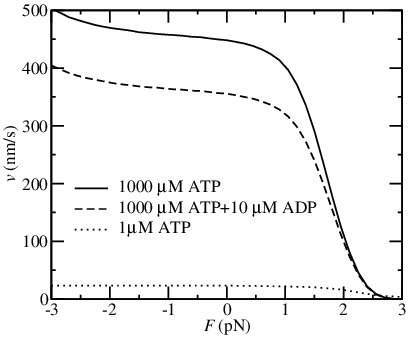}
   }
   \mycaption{Force-velocity relation of the  5-state   model with 1mM ATP
     (solid), 1mM ATP+ 10$\mu$M ADP (dashed) and 1$\mu$M ATP (dotted).  Note the
     sharper drop at high loads as compared to the  4-state  coupled model
     (Fig.~\ref{fig:9}).}
   \label{fig:10}
 \end{figure}

 \subsection*{Force-velocity and run length curves}

 The bending energies, calculated for each possible dimer configuration, and the
 transition rates were fed into a kinetic simulation to determine the average
 velocity of a dimeric motor and its dissociation rate from actin.  The most
 probable kinetic pathway of the dimer is indicated by thick arrows in
 Fig.~\ref{fig:8}, while the thin arrows indicate some of the possible side
 branches that can lead to dissociation.  Figure \ref{fig:9} shows the resulting
 force-velocity curves and Fig.~\ref{fig:11} the dissociation rates.

 An analytical solution  of the 4-state model  would, in theory, require solving
 the occupation probabilities for a system with about $6+8\times 3 \times 3=78$
 states (6 states with one head bound, plus configurations with both heads
 bound, where each head can occupy 3 different states and the relative positions
 of both heads can have 8 different values).  Such a system could easily be
 solved numerically, but would be too complex for obtaining an insightful
 analytical expression.  However, we will show that a simplified pathway can
 already lead to expressions that agree reasonably well with simulation data and
 are therefore useful for fitting model parameters to experimental data.

 In the following, we give approximate expressions for the most significant
 steps in the mechanochemical cycle.  The average time it takes for a head in
 the state 0 to bind an ATP molecule can be estimated as
 \begin{equation}
 \left< t_{\rm +ATP} \right> = \frac{1}{k_{\rm +ATP} [{\rm ATP}] }
 \left(1+ \frac{k_{\rm +ADP} [{\rm ADP}]}{k_{-\rm ADP}} \right)
 \end{equation}
 where the second term takes into account a reduction of the forward rate due to
 ADP rebinding.  The second rate limiting process (especially at hight loads) is
 the release of phosphate. The average dwell time in the state with one head
 free and the other one in the ADP.Pi state is
 \begin{equation}
   \label{eq:11}
   \left< t_{\rm -Pi} \right> = \frac {1}{k_{\rm -Pi}}
 \end{equation}
 The third rate limiting step is the ADP release, with the time constant
 \begin{equation}
 \left< t_{\rm -ADP} \right> =  \frac 1 {k_{\rm -ADP}}\;.
 \end{equation}
 With these three average dwell times, the motor velocity can be calculated as 
 \begin{equation}
   \label{eq:13}
   v=\frac{\left< d \right>}{\left< t_{\rm -Pi} \right>+ \left< t_{\rm -ADP} \right> +  
     \left< t_{\rm +ATP} \right> }\;,
 \end{equation}
 where $\left< d \right>$ denotes the average step size, which is about
 $35\,{\rm nm}$.  The individual rates that appear in this expression can be
 estimated as follows: $k_{\rm -Pi}\approx k_{\rm -Pi}^0 \exp(-F \epsilon_{\rm
   -Pi}d_{\rm PS}/k_B T)$ with $d_{\rm PS}=L (\cos \phi_{\rm ADP}-\cos \phi_{\rm
   ADP.Pi})+\delta$ and $k_{\rm -ADP}\approx k_{\rm -ADP}^0 \exp(-\Delta U_{\rm
   -ADP}/2k_B T) \approx 0.65 k_{\rm -ADP}^0 $.  The results for two different
 ATP concentrations are shown in Fig.~\ref{fig:9}A and compared with a
 simulation result.  The analytical expression reproduces the simulation result
 well, with a small deviation being mainly the result of alternative pathways,
 neglected force-dependence of the ADP release rate and variation in the step
 size.  The experimentally measured force-velocity curves
 \citep{Mehta.Cheney1999,Uemura.Ishiwata2004} are also well reproduced, although
 the experiments show a more abrupt drop in velocity at high loads, with no
 measurable effect up to about 1pN.

 In the 5-state model the power-stroke can be fast and reversible, in which case
 the pre- and the post-powerstroke state can reach an equilibrium and the
 limiting rate is proportional to the probability of the post-powerstroke state
 $1/(1+ \exp(F d_{\rm PS} /k_B T))$ - a significantly sharper load dependence
 than the 4-state model (Fig.~\ref{fig:10}).

 \subsection*{Inhibition by ADP and phosphate} 
\begin{figure}
   \figurecontents{ \includegraphics{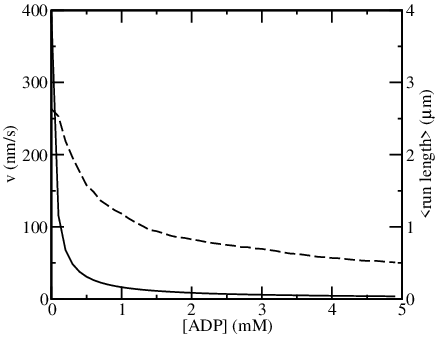} } 
   \mycaption{Velocity (continuous, left scale) and mean run length (dashed,
     right scale) as a function of ADP concentration in the 4-state model for
     zero load and 1mM ATP.}
   \label{fig:12}
 \end{figure}

 It is a well established observation that ADP can slow down myosin V by binding
 to heads in the state with no nucleotide and thereby preventing them from
 accepting an ATP molecule.  The rate of ADP rebinding is already taken into
 account in the kinetic constants and the model naturally reproduces the
 observed behavior, as shown in Figs.~\ref{fig:9} and \ref{fig:12} for the
 4-state model and in Fig.~\ref{fig:10} for the 5-state model.  Not yet
 experimentally investigated has been the inhibition by phosphate.  Its
 intensity depends on the reverse power-stroke rate, which is one of the open
 parameters of our model.  In the 4-state model, Pi re-binding is necessary for
 the reverse power stroke and therefore some inhibition effect can be expected
 at high loads.  The simulation shows clearly that the phosphate concentration
 has no effect on zero-load velocity, but it does slow down the motor close to
 stall (Fig.~\ref{fig:9}B).  A similar effect of Pi on isometric force has also
 been observed in muscle \cite{Cooke.Pate1985}.  In the 5-state model Pi
 rebinding is not mechanically sensitive and its effect is roughly
 force-independent.  However, with the parameters chosen here, it is negligible.

 \subsection*{Three dissociation pathways}  

\begin{figure}
   \figurecontents{
     \includegraphics{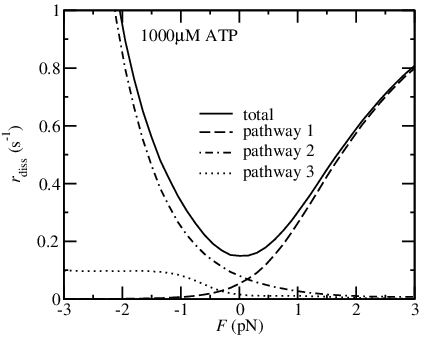}~\\
     \includegraphics{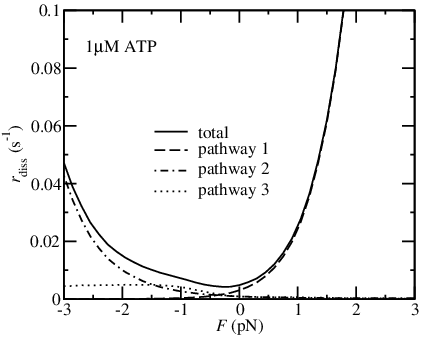}~
   }
   \mycaption{Dissociation rate of myosin V dimers from actin under a high (top)
     and a low (bottom) ATP concentration (4-state model).  The continuous line
     shows the total dissociation rate, the dashed line the dissociation via
     pathway 1, the dot-dashed line via pathway 2 and the dotted line via
     pathway 3.}
   \label{fig:11}
 \end{figure}

 As we can see from the kinetic scheme (Fig.~\ref{fig:8}), there are three
 significant pathways in the cycle that can lead to the dissociation of the
 myosin V dimer from an actin filament.  The first pathway leaves the cycle if a
 dimer bound with one head in the ADP.Pi state detaches before the second head
 can attach.  The second pathway runs through a state in which the bound head
 releases ADP and binds a new ATP molecule before the free head can bind.  With
 the third pathway we denote all processes that involve the detachment of a head
 in the ADP state.  This is the pathway favored by recent results of
 \citet{Baker.Warshaw2004}.  Figure \ref{fig:11} shows the dissociation rate,
 separated by contributions of the three pathways.  They have the following
 characteristics:

 \emph{Pathway 1:} With this pathway we denote the dissociation of a head in the
 ADP.Pi state.  Because this state is long-lived at high loads in the 4-state,
 but short-lived in the 5-state model, the resulting force-dependence of the
 dissociation rate differs significantly in both scenarios.  In the 4-state
 model, the contribution to the dissociation probability per step shows a strong
 load-dependence, but no significant dependence on the ATP concentration. It can
 be estimated as
 \begin{equation}
   \label{eq:14}
   P_{\rm diss}\approx \frac {k_{\rm -A}}{k_{\rm -Pi}}\approx \frac{k_{\rm -A}}{k^0_{\rm -Pi}}
   e^{\frac{F \epsilon_{\rm -Pi} d_{\rm PS}}{k_B T}}
 \end{equation}
 with $d_{\rm PS}=L (\cos \phi_{\rm ADP}-\cos \phi_{\rm ADP.Pi})+\delta$.  The
 dissociation rate is higher for positive loads. From the estimated run length
 at $1\,{\rm pN}$ load and saturating ATP concentration of about 15 steps
 \cite{Clemen.Rief2003}, we can estimate the unbinding rate as $k_{\rm
   -A}\approx 1\,{\rm s}^{-1}$.  In order to account for reported run lengths of
 over 50 steps at low loads, we tentatively assign $k_{\rm +A}^0\approx
 5000\,{\rm s}^{-1}$.

 In the 5-state model, the situation is reversed.  There the dissociation
 process on path 1 takes place if the trail head releases ADP before the lead
 head releases Pi, which can happen in two different ways: on one the rate is
 approximately force-independent, on the other it grows with negative (forward)
 loads.  In order to obtain a significant contribution to the detachment rate on
 this pathway, we choose a higher detachment rate $k_{-A}$ than in the 4-state
 model ($50\,{\rm s}^{-1}$ instead of $1\,{\rm s}^{-1}$).

 \emph{Pathway 2:} Because the process of unbinding requires an ATP molecule,
 the per-step dissociation rate grows with the ATP concentration.  In addition,
 it is proportional to the ratio of the $ADP$ dissociation rate and the actin
 binding rate, $k_{\rm -ADP}/k_{\rm +A}$, which is higher for negative (forward)
 loads. This holds in both the 4- and the 5-state scenario.

 \emph{Pathway 3:} The dissociation probability on pathway 3 is proportional to
 the detachment rate in the ADP state, $k'_{\rm -A}$. Of all three pathways,
 this one shows the weakest load-dependence, although it is higher for forward
 loads.

 We expect that systematic data on mean run length as a function of load and nucleotide
 concentrations will be helpful to determine the remaining model
 parameters.

 \begin{figure}[htbp]
   \figurecontents{\includegraphics{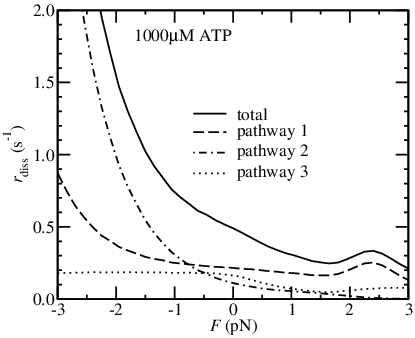}}
   \mycaption{Force-dependence of the dissociation rate in the 5-state model.
     The load dependence for positive loads is much weaker than in the 4-state
     model (Fig.~\ref{fig:11})}
   \label{fig:13}
 \end{figure}

 \subsection*{ Reverse stepping in the 5-state model }

\begin{figure}[htbp]
   \figurecontents{
     \includegraphics{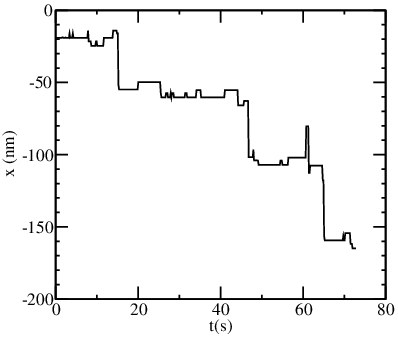}
   }
   \mycaption{Reverse stepping in the 5-state model under a high load (4.5pN),
     $10\,{\mu \rm M}$ ATP and $1\,{\mu \rm M}$ ADP.  There is also some
     creeping motion between the steps, which results from the attachment and
     detachment of the two heads on neighboring sites, and only takes place if
     myosin V is allowed to follow a helical path on actin.  If binding is
     constrained to one side of the actin filament (like on a coverslip), then
     only regular reverse steps with the periodicity of the helix are observed
     (not shown).}
  \label{fig:14}
\end{figure}

 As a consequence of both the reversibility of the power stroke and the slower
 dissociation rate at high loads, the motor can step backwards under loads
 exceeding the stall force (Fig.~\ref{fig:14}).  Note that these steps are not
 the simple reversal of forward steps (which would involve ATP synthesis), but
 rather indicate a different pathway in the kinetic scheme, in which both heads
 stay in the ADP state and alternately release actin at the leading position and
 rebind at the trailing.  The time scale of reverse stepping is determined by
 the dissociation rate of a head in the ADP state, $k'_{\rm -A}$, which we chose
 as $0.1\,{\rm s}^{-1}$.  With a higher value of $k'_{\rm -A}$, especially for
 the pre-powerstroke state (so far we assumed that the rate is equal in both ADP
 states), faster stepping would also be possible, although there is an upper
 limit on $k'_{\rm -A}$, imposed by the dissociation rate on pathway 3.

\section*{Discussion}

We used the geometrical data of the myosin V molecule as obtained from EM
images to calculate the conformations and elastic energies in all dimer
configurations.  These data were first used in a model with a four-state cycle
and subsequently in a five-state model.

The first result, which follows directly from the bending potentials and is
independent of the underlying cycle is that the elastic lever arm model
explains two key components of the coordination between heads: why the lead
head does not bind to actin before the power stroke in the trail head and why
it does not undergo its power stroke before the trail head detaches.  It also
allows us to calculate the distribution of step sizes. The results for
different lever-arm lengths (Fig.~\ref{fig:5}) give realistic values, in
agreement with step size and helicity measurements
\cite{Purcell.Sweeney2002,Ali.Ishiwata2002}, even though they have a slight
tendency towards underestimation and also show a narrower distribution than
direct electron microscopy observations \cite{Walker.Knight2000}.  A possible
explanation for the broader distribution than predicted by the model lies in
the fact that in reality the actin structure does not follow the perfect helix,
as assumed in our model, but has angular deviations of up to $10^{\circ}$ per
subunit \cite{Egelman.DeRosier1982}. Taking these fluctuations into account
would clearly broaden the distribution of our step sizes, but alone it cannot
explain the tendency towards longer steps.  The most straightforward
explanation for the longer steps is that the power stroke has an additional
right-handed azimuthal component.  Then the configuration with the lowest
energy is reached if the lead head is twisted to the right relatively to the
trail head, which is the case if it is bound further away along the helix.  The
observation that the actin repeat is often somewhat longer than 13 subunits
(some results suggest a structure closer to a 28/13 helix
\citep{Egelman.DeRosier1982}) could also partially explain the deviation.

An issue that has been much discussed is the contribution of Brownian motion
and the power stroke to the total step size.  With the geometric data used in
this study, the power stroke, i.e., the distance of the lever arm tip movement
between the states ADP.Pi and ADP, is about 31nm, or 5nm less than the average
step size.  Note that the second, smaller power stroke connected with ADP
release does not contribute to the step size because it is normally followed by
the detachment of the same head. Its function could be suppressing premature
dissociation before the lead head binds and thus improving the processivity.
The remaining 5nm can be overcome by Brownian motion before the lead head
binds.  However, at low loads, the binding of the lead head does not move the
load, but rather stores the energy into bent lever arms. This energy gets
released when the rear head detaches, which leads to an elastic power stroke
immediately preceding the power stroke upon Pi release.  At higher loads the
situation is different, because the 5nm load movement occurs when the lead head
binds.  In neither case we expect the 5nm power stroke to be resolvable under
normal conditions because it always immediately precedes or follows the large
power stroke.  However, it is possible that the substeps become observable in
the presence of chemicals that slow down the power stroke
\citep{Uemura.Ishiwata2004}.

In order to fully reproduce the substeps as reported by
\citet{Uemura.Ishiwata2004}, some modifications would be necessary to the
model.  First, part of the power-stroke would have to occur immediately upon Pi
release, resulting in a lever arm move of about $12\,\rm nm$ (first substep).
This step would need a very strong force-dependence in its transition rate
(activation point near the final state).  The subsequent longer power stroke
(ADP'$\to$ADP) would then need a slower rate ($\sim 200\,\rm s^{-1}$) with less
force dependence (activation point close to the initial state).  However, the
finding that the substep position is independent of force remains difficult to
explain, because the substep involves transition between a stiff configuration,
bound on both heads, and a more compliant state, bound on a single head.  

The main value of both models (4- and 5-state) is that they provide a
quantitative explanation of the coordinated head-over-head motility of the
dimeric molecule while using only the properties of a single head as input.
Both models also explain the observed force-velocity curves at high and low ATP
concentration and the effect of additional ADP, but these features already
reveal some testable differences between the two scenarios.  One of them is the
shape of the force-velocity curve.  In the  4-state  scenario the reverse
power-stroke needs the rebinding of a phosphate molecule.  This makes the
cutoff behavior at high loads dependent on the Pi concentration: the velocity
drop is more gradual at low, but might become sharper at high Pi concentrations
(Fig.~\ref{fig:9}B).  In the 5-state scenario the velocity decline
is more abrupt regardless of the Pi concentration.  This is the first
suggestion how experiments with improved precision and a wider range of
chemical conditions could help distinguishing between the two scenarios.

The main difference between the two scenarios is the predicted shape of the run
length. Because the dissociation can take place on three different pathways,
its rate depends on a number of parameters, of which a few cannot yet be
determined by other methods.  In the  4-state   model the dissociation
rate at high loads is dominated by detachment of a head in the ADP.Pi state and
it therefore depends on the ratio $k^0_{\rm -A}/ k^0_{\rm -Pi}$
(Eq.~\ref{eq:14}). A strong increase with the load is characteristic
for the  4-state   model, because the load slows down the phosphate
release and prolongs the dwell time in the state that is most vulnerable to
dissociation.  Dissociation at negative (forward) loads is dominated by
pathways 2 (ATP mediated actin release in one head before the other head has
bound) and 3 (dissociation of a head with ADP).  In the  5-state  model
all three pathways can contribute towards the dissociation rate, but there is
no significant increase for positive loads - in fact, the dissociation rate can
even decrease.

The run length shortens with an increasing ADP concentration in both scenarios.
The decrease in run length is weaker than the decrease in the velocity
(Fig.~\ref{fig:12}), which is consistent with recent observations
\cite{Baker.Warshaw2004}.  However, we cannot reproduce the reported complete
saturation of run length at hight ADP concentrations.
\citet{Baker.Warshaw2004} explain this saturation with a big difference
(50-fold) between the attachment rates of the lead head depending whether the
trail head is in the ADP or apo state, which we currently cannot reproduce with
the relatively small power stroke ($10^{\circ}$) upon ADP release in our model.

An interesting difference between the 4- and the 5-state model is also that the
5-state model allows backward steps at high loads (above the stall force),
while the 4-state model predicts rapid dissociation.  In general, there are
three possibilities how backward steps can occur: (i) The motor hydrolyzes ATP,
but runs backwards.  (ii) The motor slips backwards without hydrolyzing ATP ---
this is the case in our model. (iii) The motor synthesizes ATP from ADP and
phosphate while being pulled backwards, as assumed by tightly coupled
stochastic stepper models \citep[e.g.,][]{Kolomeisky.Fisher2003}.  It is
possible to test these three possibilities experimentally: If (i) is the case,
the backward sliding velocity should show a Michealis-Menten type dependence on
ATP concentration.  This mechanism would, however, require an even looser
mechanochemical coupling, so that not only the release of Pi, but also the
release of ADP and binding of ATP would be possible without completing the
power stroke.  In case (iii) it should depend on ADP as well as on Pi
concentration, but not on ATP.  In case (ii), which is favored by our study,
the backward stepping occurs when both heads have ADP bound on them and they
successively release actin at the lead position and rebind it at the new trail
position.  Even though this stepping requires no net reaction between the
nucleotides, a certain (low) ADP concentration is still required to prevent the
heads from staying locked in the rigor (no nucleotide) state.

The application of the elastic lever-arm approach developed here should not be
limited to simple geometries and longitudinal loads.  A natural extension of
the present work will be the influence of perpendicular forces on the activity
of the motor.  One will also be able to study the stepping behavior in more
complex geometries, for example when passing a branching site induced by the
Arp2/3 complex \cite{Machesky.Gould1999}.

After completion of this manuscript, it has been brought to my attention that
\citet{Lan.Sun2005} have also published a model for myosin V, based on the
elasticity of the lever arm. In contrast to our model, they do not describe it
as an isotropic rod, but use a weaker in-plane stiffness, combined with a
strong (phenomenological) azimuthal term that prevents binding of both heads to
adjacent sites on actin.  Another difference is that their study explicitly
excludes dissociation events, whereas we use the dissociation rate to determine
some of the model parameters.

\section*{Acknowledgment}
I would like to thank Erwin Frey and Jaime Santos for help with calculating the
lever-arm shape, Peter Knight for help with the geometry of the molecule, and
Matthias Rief and Mojca Vilfan for helpful discussions.  This work was
supported by the Slovenian Office of Science (Grants No.~Z1-4509-0106-02 and
P0-0524-0106).

\section*{Appendix}
\subsection*{Numerical solution for the lever arm shape}

The aim of this calculation is to determine the shape of the dimeric
molecule for a given set of binding sites (trailing head bound on the
site with the index $i_1$, leading head with $i_2$), nucleotide states,
which determine the lever arm  starting  angles $\phi_1$ and $\phi_2$, and a
given external load $F$.

We start this task by deriving a function that numerically determines the
endpoint of a lever arm as a function of the force acting on it:
$\mathbf{x}_{j}(\mathbf{F}_{j}, \phi_{j})$ ($j=1,2$).  The shape of the whole
molecule can then be determined numerically from the conditions that the
endpoints of the two lever arms coincide, $\mathbf{x}_1=\mathbf{x}_2$, and from
the force equilibrium in that point
\begin{equation}
\label{eq:15}
\mathbf{F}_1+\mathbf{F}_2 = -F \hat{e}_x\;.
\end{equation}
In many cases the function $\mathbf{x}_{j}$ will have more than one solution.
Then we solve the system with all possible combinations and then choose the
solution with the lowest energy $U=U_1+U_2+Fx$, where $U_1$ and $U_2$ denote
the energy stored in the distortion of each lever arm and $Fx$ the work
performed against the applied load.

For a head bound at site $i$, the position of the proximal end of its lever arm
in Cartesian coordinates reads
\begin{equation}
  \label{eq:16}
  \mathbf{x}^0=\left( \begin{array}{c}
      i a+\delta\\
      - R \sin(\theta) \\
      R \cos(\theta)
    \end{array}
  \right)
\end{equation}
and its initial tangent
\begin{equation}
  \label{eq:17}
  \hat{t}^0=\left( \begin{array}{c}
      \cos(\phi) \\
      - \sin(\phi) \sin(\theta) \\
      \sin(\phi) \cos(\theta)
    \end{array}
  \right)
\end{equation}
where $\phi$ is the lever arm tilt (a function of the nucleotide state),
$\delta$ is the relative position of the lever arm proximal end ($0$ or
$3.5\,{\rm nm}$) and $\theta$ is the azimuthal angle of the actin subunit to
which the head is bound, $\theta=\theta_0 i$ with $\theta_0\approx \frac{6}{13}
\times 360^{\circ} \approx 166^{\circ}$.  The helix rise per subunit is $a=
2.75\,{\rm nm}$.

If the force $\mathbf{F}$ acts on a lever arm that leaves the head in the
direction $\hat{t}^0$, the whole lever arm will be bent in a plane spanned by
the vectors $\hat{t}^0$ and $\mathbf{F}$.  We can introduce a new
two-dimensional orthogonal coordinate system in this plane, so that
\begin{align}
  \tilde{\hat{t}}^0&=\left( \begin{array}{c}0\\1\end{array} \right) &
  \tilde{\mathbf{F}}&=\left( \begin{array}{c}\tilde F _x \\ \tilde F _y
    \end{array}\right) \\ \tilde F_y&=\mathbf{F} \hat{t}_0 & \tilde F_x&=\left|
    \mathbf{F}-\hat{t}_0 ( \mathbf{F} \hat{t}_0 ) \right|
\end{align}
In this coordinate system the shape can be determined by solving the equations
\begin{align}
\label{eq:20}
M(s)&=\tilde{\mathbf{F}} \wedge ( \tilde{\mathbf{x}}(L)-\tilde{\mathbf{x}}(s)
)= EI \frac{d\phi(s)}{ds} \\ \frac{d\tilde{\mathbf{x}}}{ds} &= \hat{\tilde{t}}
\qquad \hat{\tilde{t}}=\left( \begin{array}{c} \sin(\phi) \\ \cos(\phi)
  \end{array} \right)
\end{align}
with the boundary condition $\phi(0)=0$.  The symbol ``$\wedge$'' denotes the
outer product, which is the out-of-plane component of the vector product. If we
differentiate Eq.~\ref{eq:20} by $\phi$ we get
\begin{equation}
  \label{eq:22}
  EI \frac{d^2 \phi}{ds^2} =  -  \tilde{F}_x \cos(\phi) +  \tilde{F}_y \sin(\phi)
\end{equation}
Through partial integration and taking into account the boundary condition
$M(L)=0$, we finally obtain
\begin{equation}
  \label{eq:23}
  \begin{split}
    &\frac{EI}{2}\left( \frac{d\phi}{ds} \right)^2 \\&= \tilde{F}_x
    (\sin\phi_L-\sin\phi)+\tilde{F}_y (\cos \phi_L - \cos \phi )\\
    &\equiv F \sin\left( \frac{\phi_L-\phi}{2}\right) \sin \left( \phi_F
      -\frac{\phi_L+\phi}{2} \right)
\end{split}
\end{equation}
Here we introduced the force angle $\phi_F$, so that $\tilde{F}_x=F
\sin(\phi_F)$ and $\tilde{F}_y=F \cos(\phi_F)$.

\begin{figure}[tbp]
  \figurecontents{ \includegraphics{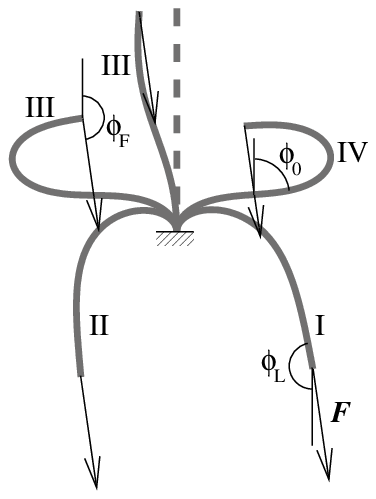} } 
  \mycaption{ The shapes of an elastic beam anchored at one end and pulled by a
    given force $\mathbf F$ on its other end.  The dashed line shows the
    unloaded beam. According to the sign of the initial curvature and the final
    angle $\phi_L$ the solutions can be divided into 4 classes.  The beam
    corresponds to the myosin V lever arm, which is anchored in the head at one
    end and connected to a flexible joint at the other end.  Note that the
    bending shown is exaggerated in comparison with realistic dimer
    configurations.  }
  \label{fig:15}
\end{figure}

Because of the ambiguity of a quadratic equation, Eq.~\ref{eq:23} generally has
two solutions for a given set of values for $\phi(s)$, $F$, $\phi_L$ and
$\phi_F$.  As we have defined the coordinate system in a way that $\tilde{F}_x
\ge 0$, we have $0 \le \phi_F \le \pi$.  We also restrict ourselves to
solutions with $\left| \phi(s) \right|<2\pi$, i.e., we do not consider any
spiraling solutions, because they always have a higher bending energy than the
straighter solution with the same endpoint.  There are four classes of
functions $\phi(s)$ that satisfy the condition that the RHS of Eq.~\ref{eq:23}
be positive:

\begin{center}
\begin{tabular}{cccc}
Solution& $\phi_L$ & $\phi(s \to 0)$ & conditions  \\
\hline
I & + & + & $0 \le \phi_L \le \phi_F$ \\
II & - & - &  $\phi_F-2\pi \le \phi_L \le  2\phi_F-2\pi$   \\
III & + & - & $0\le \phi_L \le \phi_F$ \\
IV & - & + &  $\phi_F-2\pi \le \phi_L \le  2\phi_F-2\pi$   \\
\end{tabular}
\end{center}

The solutions III and IV have a turning point at $\phi_0=-2
(\pi-\phi_F) -\phi_L$, where $d\phi/ds$ changes sign.
Eq.~\ref{eq:23} can finally be transformed to
\begin{align}
  \label{eq:24}
  L&=\frac{1}{2} \sqrt{\frac{EI}{F}} I(\phi_L) \qquad (\text{cases I and II})\\
  L&=\frac{1}{2} \sqrt{\frac{EI}{F}} (2 I(\phi_0)+I(\phi_L))  \qquad (\text{cases III and IV})\\
I(\phi_x)&=\left| \int_0^{\phi_x} \left(  \sin\left(
  \frac{\phi_L-\phi}{2}\right) \sin \left( \phi_F -\frac{\phi_L+\phi}{2}
  \right) \right) ^{-1/2} d\phi \right| \nonumber 
\end{align}
Note that for classes II and III the RHS of Eq.~\ref{eq:24} is not monotonous
in $\phi_L$ and there can be two solutions for a given $L$.  Taking this into
account, we obtain a total of up to 6 solutions.  A situation in which all
cases are represented is shown in Fig.~\ref{fig:15}.

The configuration of the dimer is determined by solving Eq.~\ref{eq:15} for all
possible combinations of modes and taking the one with the lowest potential.
The numerical integration and solution were performed using NAG libraries
(Numerical Algorithms Group) and the 3-d graphical representation of the
calculated shapes was made with POV-Ray (www.povray.org).

\end{document}